\documentclass[aps,prl,preprint,citeautoscript,superscriptaddress,showpacs]{revtex4-1}

\usepackage{graphicx}
\usepackage{amsmath}
\usepackage{hyperref}
\usepackage{epstopdf}

\begin{document}


\title{Charge separation by photoexcitation in semicrystalline polymeric semiconductors: An intrinsic or extrinsic mechanism?}



\author{Francis~Paquin}
\affiliation{D\'{e}partement de physique \& Regroupement qu\'{e}b\'{e}cois sur les mat\'{e}riaux de pointe, Universit\'{e} de Montr\'{e}al,  C.P.\ 6128, Succursale centre-ville, Montr\'{e}al (Qu\'{e}bec), H3C~3J7, Canada}

\author{Gianluca~Latini}
\altaffiliation{Current address: Centre for Biomolecular Nanotechnologies, Italian Institute of Technology, Lecce, Italy.}
\affiliation{Department of Materials and Centre for Plastic Electronics, Imperial College London, South Kensington Campus, London SW7~2AZ, United Kingdom}

\author{Maciej~Sakowicz}
\affiliation{D\'{e}partement de physique \& Regroupement qu\'{e}b\'{e}cois sur les mat\'{e}riaux de pointe, Universit\'{e} de Montr\'{e}al,  C.P.\ 6128, Succursale centre-ville, Montr\'{e}al (Qu\'{e}bec), H3C~3J7, Canada}

\author{Paul-Ludovic~Karsenti}
\affiliation{D\'{e}partement de physique \& Regroupement qu\'{e}b\'{e}cois sur les mat\'{e}riaux de pointe, Universit\'{e} de Montr\'{e}al,  C.P.\ 6128, Succursale centre-ville, Montr\'{e}al (Qu\'{e}bec), H3C~3J7, Canada}

\author{Linjun~Wang}
\affiliation{Chemistry of Novel Materials, University of Mons, Place du Parc 20, B-7000 Mons, Belgium}

\author{David~Beljonne}
\affiliation{Chemistry of Novel Materials, University of Mons, Place du Parc 20, B-7000 Mons, Belgium}

\author{Natalie~Stingelin}
\affiliation{Department of Materials and Centre for Plastic Electronics, Imperial College London, South Kensington Campus, London SW7~2AZ, United Kingdom}

\author{Carlos~Silva}
\email[E-mail: ]{carlos.silva@umontreal.ca}
\affiliation{D\'{e}partement de physique \& Regroupement qu\'{e}b\'{e}cois sur les mat\'{e}riaux de pointe, Universit\'{e} de Montr\'{e}al,  C.P.\ 6128, Succursale centre-ville, Montr\'{e}al (Qu\'{e}bec), H3C~3J7, Canada}


\date{\today}

\begin{abstract}
We probe charge photogeneration and subsequent recombination dynamics in neat regioregular poly(3-hexylthiophene) films over six decades in time by means of time-resolved photoluminescence spectroscopy. Exciton dissociation at 10K occurs extrinsically at interfaces between molecularly ordered and disordered domains. Polaron pairs thus produced recombine by tunnelling with distributed rates governed by the distribution of electron-hole radii. Quantum-chemical calculations suggest that hot-exciton dissociation at such interfaces results from a high charge-transfer character.

\end{abstract}


\maketitle


Unravelling primary electronic processes in polymeric semiconductors opens a fundamental window to their materials physics, which is of central importance for emerging applications in optoelectronics. The steps to generate charge by optical absorption are currently the subject of wide interest~\cite{*[{See }][{ for a comprehensive review.}] Clarke2010}.  Here, we focus on charge generation and recombination dynamics in \emph{neat} regioregular poly(3-hexylthiophene) (P3HT). This semicrystalline polymer adopts $\pi$-stacked lamellar microstructures in the solid state~\cite{Sirringhaus:1999p185}, leading to two-dimensional electronic dispersion~\cite{Spano:2006p742}. Crystallinity induced by molecular organization influences profoundly electronic properties, exemplified by 
the high yield ($\eta$) of apparently direct charge photogeneration. Whereas in general $\eta \ll1$\% in less organized polymeric semiconductors~\cite{Silva:2002p43}, various groups have reported $ \eta$ up to 30\% over ultrafast timescales in P3HT films at 300\,K~\cite{Hendry:2004p1534,Ai:2006p6883,Sheng:2007p4477,Cunningham:2008p6673,Piris:2009p4569}. This is very surprising because photoemission spectroscopy measurements place the energy for free polaron generation $\sim0.7$\,eV above the optical gap in neat regioregular P3HT~\cite{Deibel:2010p5883}. Weak interchain electronic coupling in the lamellar architecture leads to a free-exciton bandwidth --- the pure electronic bandwidth due to dispersion neglecting coupling to vibrations --- that is well below this energy offset in P3HT stacks~\cite{Spano:2005p740,*Spano:2007p744,Clark:2007p51,Spano:2009p28}. 
Thus, are charges indeed generated \emph{intrinsically} (directly by photoexcitation, resulting from the crystalline electronic structure) or \emph{extrinsically} (due to a driving force for dissociation following exciton formation)?

Charge photogeneration in neat P3HT has been studied previously by means of transient absorption spectroscopy~\cite{Sheng:2007p4477,Piris:2009p4569}, which measures directly the dynamics of nascent polarons. In this letter, we implement time-resolved photoluminescence (PL) spectroscopy at 10\,K to probe charge photogeneration and recombination dynamics. Our strategy is to exploit the intricate detail of electronic structure, structural relaxation, and correlated disorder afforded by the evolution of the spectral bandshape of the PL spectrum~\cite{Spano:2009p28}, over timescales where it is known that charge photogeneration is important. By analyzing the bandshape and decay dynamics of delayed PL due to charge-pair recombination, we probe the recombination environment, which also probes the environment in which photogeneration occurs, since photoexcitations are frozen at 10\,K. 
We find that charge generation occurs over sub-nanosecond timescales by dissociation of excitons created at interfaces between lamellar (aggregate) and poorly stacked (non-aggregate)  domains, driven by energetic disorder. Thus, prompt charge photogeneration is an \emph{extrinsic} process, and film microstructure determines the surface area and the energy landscape of interfaces between domains. 

PL measurements were carried out with a 40-fs, 532-nm (2.33-eV) pulse train derived from an optical parametric amplifier (Light Conversion TOPAS), pumped by a Ti:sapphire laser system (KMLabs Dragon). Time-resolved PL spectra were measured with an optically-triggered streak camera (Axis-Photonique, 6-ps instrument response). Alternatively, an intensified CCD camera (Princeton-Instruments PIMAX 1024HB) was used. P3HT films (Merck, $M_w = 47.8$\,kg/mol, $M_n = 26.2$\,kg/mol, polydispersity = 1.83, 150-nm thick) were spun from trichlorobenzene solution (6\% wt).

In P3HT films, the PL spectrum is understood within the framework of a weakly-coupled H-aggregate model~\cite{Clark:2007p51}, resulting from weak resonance-Coulomb coupling ($J$) of transition moments aligned co-facially in neighboring polymer chains~\cite{Spano:2005p740,*Spano:2007p744}. In contrast, the absorption spectrum contains contributions from both the aggregate and non-aggregate regions~\cite{Clark:2007p51}. From the ratio of the (0,0) and (0,1) absorbance peaks~\cite{EPAPS}, we estimate a free-exciton bandwidth ($W = 4J$) of $100 \pm 3$\,meV~\cite{Clark:2007p51}. Furthermore, using a Franck-Condon model for the absorption spectrum of the H aggregate~\cite{Clark:2009p3235}, we estimate a volume fraction $52 \pm 5$\% for aggregates (electronically coupled chains) versus non-aggregate regions.


 \begin{figure}
 \includegraphics[width=75mm]{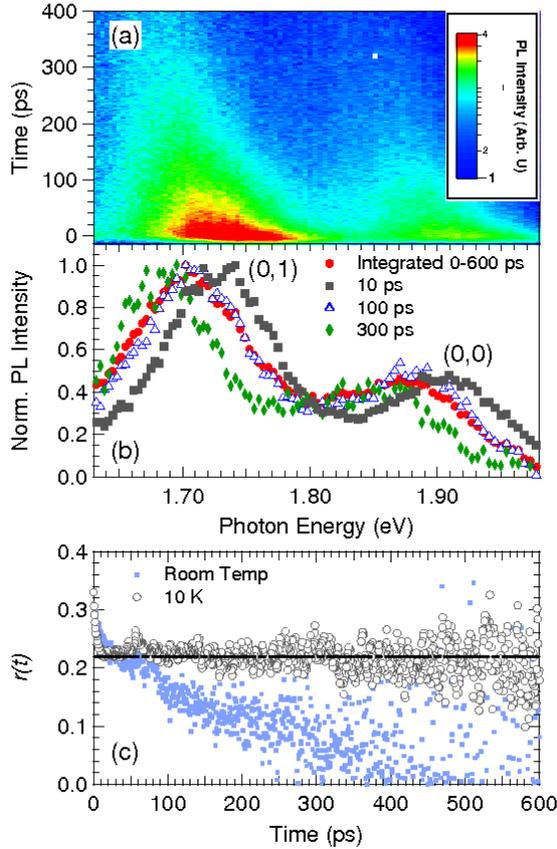}%
 \caption{(Color online) (a) Time-resolved PL spectrum, exciting at 2.33\,eV with fluence 19\,$\mu$J\,cm$^{-2}$, measured at 10\,K. (b) Normalized slices of (a) at times indicated in the legend. (c) Time-resolved PL anisotropy $r$ at 10\,K (open circles) and at room temperature (dots). The horizontal line indicates the average anisotropy, $\bar{r} = 0.22 \pm 0.03$, between 20 and 600\,ps.
 \label{fig:streak}}
 \end{figure}

We carried out time-resolved PL measurements at 10\,K to explore early-time exciton dynamics in neat P3HT, shown in Fig.~\ref{fig:streak}(a). The PL intensity decays substantially over sub-nanosecond timescales, and the spectrum red-shifts by  $> 40$\,meV with weak evolution of the spectral bandshape. We examine in more detail this spectral evolution in Fig.~\ref{fig:streak}(b). Fig.~\ref{fig:streak}(c) displays the early-time PL anisotropy, defined as $r =  (I_{\|} - I_{\bot}) / (I_{\|} + 2I_{\bot})$, where $I_{\|(\bot)}$ is the instantaneous PL intensity parallel (perpendicular) to the linear polarization of the pump pulse. At 10\,K (open circles), $r$ decays to 0.23 within the instrument response of several picoseconds due to exciton self trapping in the photophysical aggregate~\cite{Chang:2007p6389,Collini:2009p3772}, and does not evolve further on a sub-nanosecond window. In contrast, $r$ decays to zero after 500\,ps at room temperature. PL anisotropy decay is a signature of exciton diffusion~\cite{[{We excite both H-aggregates and non-aggregates with similar probability at 2.33\,eV~\cite{EPAPS}, so exciton diffusion would result in a dynamic memory loss of the ensemble-average transition-dipole-moment orientation. We observe this at 300\,K because exciton diffusion is thermally activated, but not at 10\,K. See }][{ for relevant modelling.}]Beljonne2005}, so at 10\,K, excitons are immobile over the timescale of the dynamic red shift in Fig.~\ref{fig:streak}(a). 
Another possibility could be relaxation of excess vibrational energy following ultrafast excitation of the sample. Parkinson~et~al.\ have reported that torsional relaxation of the backbone leads to more cofacial lamellar structures (more extended correlation of site energies in the aggregate~\cite{Spano:2009p28}), leading to a dynamic loss of (0,0) intensity over $\sim 13$\,ps at 300\,K~\cite{Parkinson:2010p6943}. We do not observe significant relaxation of the (0,0) relative intensity over the timescale of Fig~\ref{fig:streak}(a). Furthermore, upon increasing the temperature even slightly, the steady-state PL spectrum broadens significantly at 10\,K~\cite{Clark:2007p51}, so we rule out thermal relaxation. 
We suggest that the dynamic spectral red-shift of the PL spectrum results from an evolving electric field distribution~\cite{kersting} from photogenerated charges in the aggregate, which is consistent with charge photogeneration yields reported in the literature and our own transient absorption measurements~\cite{EPAPS}. If so, charges are generated over all timescales spanning $\lesssim1$\,ns, by a mechanism not involving exciton diffusion.

 \begin{figure}
 \includegraphics[width=86mm]{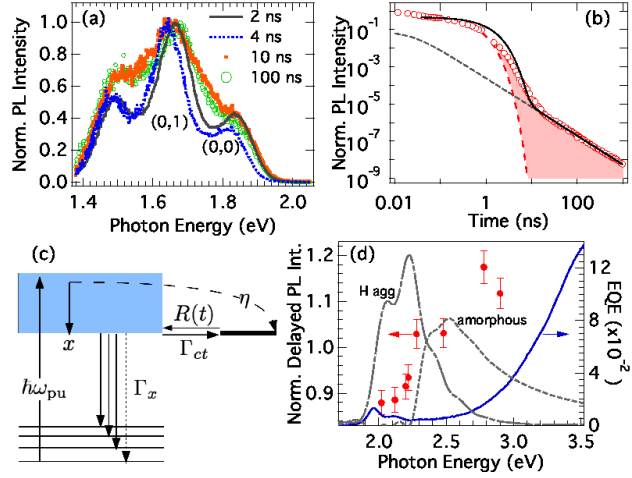}%
 \caption{(Color online) (a) Time-resolved PL spectra measured after 2.33-eV excitation at 10\,K, and at gate times denoted in the legend. (b) Spectrally integrated PL intensity versus gate delay. The time-resolved PL intensity from Fig.~\ref{fig:streak} is included over sub-nanosecond timescales. The dashed lines represent a two-exponential fit with $\langle \tau \rangle = 235 \pm 23$\,ps and an asympototic power law [$1/(1 + (t/\tau^{\prime})^{\alpha})$] with $\alpha = 1.54 \pm 0.08$ and $\tau^{\prime} \ll \langle \tau \rangle$. A fit to the model (eq.~\ref{eq:PLmodel}), represented schematically in (c), is shown as a continuous line (for the case of $\eta = 0.03$, $\Gamma_{ct}= 0.35$\,ns$^{-1}$, $\Gamma_x = 1.23$\,ns$^{-1}$, $\nu = 4.37 \times 10^{13} $\,s$^{-1}$,  $\mu = 0.54$). (d) Delayed PL excitation spectrum at 10\,K, measured by phase-sensitive methods as described in auxiliary information~\cite{EPAPS}, superimposed on the components of the H-aggregates and non-aggregates of the absorption spectrum (see ref.~\citenum{Clark:2009p3235}). Also shown is the external quantum efficiency spectrum measured at 300\,K in a ITO/PEDOT:PSS/P3HT/LiF/Al diode fabricated and encapsulated in a N$_2$-filled glove box.
  \label{fig:ICCD}}
 \end{figure}

We next consider delayed PL dynamics due to charge recombination. Fig.~\ref{fig:ICCD}(a) displays time-gated PL spectra at 10\,K. After a few nanoseconds,  the aggregate spectrum is superimposed with a broader component (see ref~\cite{EPAPS}) with similar dynamics to those of the H aggregate since this composite bandshape persists over microseconds. While we know that the aggregate spectrum arises from lamellar stacks, we conjecture that the featureless spectrum arises from non-aggregate regions. It is reminiscent of red-emitting species found in amorphous polymer films, often referred to as excimers~\cite{Schwartz:2003p1529}. From the delayed PL bandshape in Fig.~\ref{fig:ICCD}(a), we conclude that recombination occurs at the interface between H-aggregate and non-aggregate regions, and can populate either species.

The total PL intensity decays exponentially over picosecond timescales, but asymptotically as a power law ($I(t) \propto t^{-1.54}$) for times much longer than the exciton lifetime (Fig.~\ref{fig:ICCD}(b)). This behavior can arise from unimolecular charge recombination~\cite{Schweitzer:1999p6711} or by triplet bimolecular annihilation~\cite{Gerhard:2002p6693,Rothe:2005p1459}. We find that the delayed PL intensity varies linearly with the pump fluence~\cite{EPAPS}, which permits us to assign the power-law decay to exciton regeneration by charge recombination with a distribution of rates, as the other two possibilities would lead to a non-linear fluence dependence. The power-law decay is independent of temperature~\cite{EPAPS}, suggesting recombination by tunnelling. We note that we cannot reproduce the measured time-resolved PL intensity as a simple superposition of a multiexponential and an asymptotic power-law decay, as the time window spanning 1--10\,ns features a PL decay over nearly three orders of magnitude that deviates significantly from either decay function (Fig.~\ref{fig:ICCD}(b)). This indicates that the two PL decay phenomena are not independent, but that one decay regime evolves into the other, with competing kinetics on nanosecond timescales. By integrating the PL intensity over timescales where the decay is non-exponential, we find that $\geq 12$\% of the time-integrated intensity is accounted for by slow recombination. This reflects a significant density of charge pairs.

Based on the results of Figs.~\ref{fig:streak} and \ref{fig:ICCD} and building upon previous reports~\cite{Hendry:2004p1534,Ai:2006p6883,Sheng:2007p4477,Cunningham:2008p6673,Piris:2009p4569
}, we construct the following photophysical picture, depicted schematically in Fig.~\ref{fig:ICCD}(c). Upon photoexcitation, charge pairs are generated with efficiency $\eta$, and the rest of the population relaxes to self-trapped excitons $x$ in $<$1\,ps~\cite{Collini:2009p3772}. These decay to the ground state with rate constant $\Gamma_x$, or charge-separate with rate constant $\Gamma_{ct}$. Charge pairs may recombine to regenerate $x$ with a temporal distribution $R(t)$. 
Thus,  
\begin{equation}
\frac{\mathrm{d}x}{\mathrm{d}t} = -(\Gamma_{x} + \Gamma_{ct})x + \eta R(t) + \int_{0}^{t}  \Gamma_{ct} x(t^{\prime}) R(t - t^{\prime})\, \textrm{d}t^{\prime}.
\label{eq:PLmodel}
\end{equation}
Assume a distribution of charge-pair distances $f(r) = \epsilon e^{-\epsilon r}$. The tunnelling-rate distance dependence is $k(r) = \nu e^{-\beta r}$ and the tunnelling-time distribution is $R(t) = \int_0^{\infty} f(r) k(r)  e^{ -k(r) t }\,dr$. With $x(0) = 1- \eta$, the Laplace transform of eq.~\ref{eq:PLmodel} is  
\begin{equation}
\hat{x}(s) = \frac{1- \eta + \eta \hat{R}(s)}{s + \Gamma_x + \Gamma_{ct} \left[ 1 -  \hat{R}(s)\right ] },
\label{eq:laplace}
\end{equation}
where $\hat{R}(s) = \mu \int_0^{\infty} e^{-(1 + \mu)\chi} / (s/\nu + e^{-\chi})\, d\chi$, $\chi = \beta r$, and $\mu = \epsilon / \beta$. The model predicts biphasic dynamics with PL intensity,  $I(t) \propto \Gamma_x x(t)$, evolving as $I(t) \propto \exp(-\Gamma_x t)$ at short times and as $I(t) \propto t^{-(1 + \mu)}$ at long times. We evaluate eq.~\ref{eq:laplace} as described elsewhere to recover $I(t)$~\cite{Brosseau:2010p5555}.

The most robust parameter in this model is the ratio of the characteristic electron-hole separation ($\epsilon$) and the distance dependence of charge tunnelling ($\beta$): $\mu = \epsilon / \beta = 0.54 \pm 0.08$, as it defines uniquely the slope of the power-law decay at long times. Hence, $\epsilon \approx 0.54 \beta$; a significant fraction of charge pairs do not recombine in the microsecond timescale, as the tail of $f(r)$ extends beyond the characteristic lengthscale of $k(r)$. The charge-pair population is expected to survive on much longer timescales, which is consistent with reports of a high residual charge density in P3HT at steady state~\cite{Dicker:2005p6940,Ferguson:2008p6925}.

We next consider $\eta$. We cannot extract it uniquely from the model, as it is coupled to $\Gamma_{ct}$. Both parameters affect the amplitude of the power-law decay without altering the decay rate. We probe the range of $\eta$ that is consistent with the literature, and explore the consequences on $\Gamma_{ct}$. With $\Gamma_x = 1.23$\,ns$^{-1}$ (fixed by the slow part of the bi-exponential decay in Fig.~\ref{fig:streak}(a)) and with $\nu = 4.37 \times 10^{13} $\,s$^{-1}$ (corresponding to the frequency of the aromatic C---C stretch measured by Raman spectroscopy~\cite{Gao:2010p5763}), we find that if we set $\eta = 3$\%, we require $\Gamma_{ct} = 0.35 \pm 0.05$\,ns$^{-1}$, but if $\eta = 30$\% we can reproduce the data with $\Gamma_{ct} = 0$. With $\eta = 40$\%, we can no longer obtain a satisfactory fit of the amplitude of the power-law component, which is over-estimated. We have interpreted the dynamic red shift of the PL spectrum (Fig.~\ref{fig:streak}) as a consequence of an evolving electric field on the subnanosecond timescale, which is comparable to the exciton lifetime. In order to rationalize our time-resolved spectroscopic data, we therefore consider that $\Gamma_{ct} \lesssim \Gamma_x$, which from the modelling, would imply that $\eta < 10$\% \emph{in the solid-state microstructure of our films}.

 \begin{figure}
 \includegraphics[width=86mm]{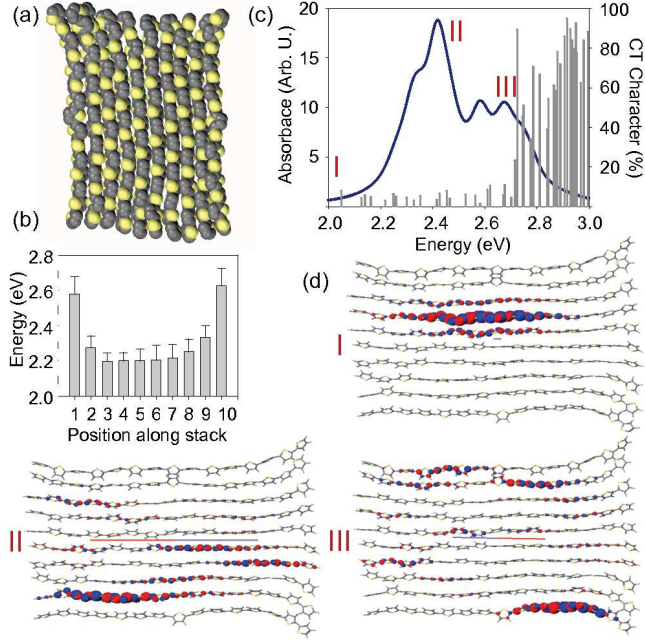}%
 \caption{(Color online) Results of quantum-chemical calculations~\cite{QuantumChem}. (a) Snapshot of the architecture of the aggregate (15-mers in a stack of 10 chains). (b) Average excitation energy (error bar is the variance) for the (10) chains in the stack, averaged over a large number of excited-state calculations on isolated conformers extracted from the stack. (c) Absorption spectrum for one configuration of the full stack superimposed to the corresponding charge-transfer character. (d) Transition density diagrams (ground-state -- excited-state overlap) for three excitations of increasing energy.
  \label{fig:quantumchem}}
 \end{figure}

If the two slow-decaying emissive species in Fig.~\ref{fig:ICCD}(a) are due to regeneration of aggregate and non-aggregate excitations by charge tunnelling, we propose that charges are initially generated at the interface between the two domains. To explore this further, we have carried out quantum chemical calculations~\cite{QuantumChem} on a stack of ten oligothiophene chains (Fig.~\ref{fig:quantumchem}(a)) to represent the crystalline moieties in the higher  molecular-weight (i.e.\ longer chain) P3HT used in our experimental study. We note higher conformational disorder at the top/bottom of the stack, leading to a higher excitation energy in those regions compared to the center (Fig.~\ref{fig:quantumchem}(b)). Fig.~\ref{fig:quantumchem}(c) displays the calculated absorption spectrum in a given configuration. For all stack configurations studied, we always observe a shoulder at $\sim2.7$\,eV on the blue side of the main absorption band at $\sim2.4$\,eV. (Note that these energies are overestimated, but the $\sim0.3$\,eV difference is meaningful.) The excited states that are generated in this spectral region are quasi-degenerate with the lowest-lying charge-transfer (CT) states, shown by the superimposed plot of CT character as a function of excitation energy. We can identify three regimes of the spatial distribution of transition densities, shown in Fig.~\ref{fig:quantumchem}(d). The excitation with the lowest transition energy (I) is always confined to the center of the stack over two to three sites as a result of disorder and is weakly emissive (H aggregate); the intermediate excitation (II) carries most of the oscillator strength and is delocalized in the center of the stack; the higher-lying excitation (III) is dominated by conformationally disordered chains at the ends of the stack, but it communicates with chains in the center so that charge separation of these higher-energy states is possible. In our measurements, we excite $\sim0.3$\,eV above the (0,0) absorption, placing us at the limit of CT excitation predicted by Fig.~\ref{fig:quantumchem}(c). We have measured the delayed PL excitation spectrum, shown in Fig~\ref{fig:ICCD}(d). This reveals that excitation on the edge of the component of the absorption spectrum due to less ordered or amorphous species enhances the delayed PL yield, consistent with the significant CT character of excitons at these photon energies. Photocurrent measurements reveal that a further 0.4-eV energy is required to produce charge carriers (Fig.~\ref{fig:ICCD}(d)), underlining that our delayed PL measurements probe recombination of tightly-bound geminate polaron pairs (not carriers) produced by excitation with energies $\lesssim 3$\,eV. Deibel~et~al.\ reported that 0.42-eV excess energy above the optical gap is required to produce polaron pairs, and then a further 0.3\,eV to generate photocarriers~\cite{Deibel:2010p5883}, consistent with our findings.  

We have presented compelling evidence that excitons with sufficient energy dissociate \emph{extrinsically} near interfaces between molecularly organized and less organized domains, underlining the key role of the disordered energy landscape in semicrystalline microstructures. This points to the importance of hot-exciton dissociation in \emph{neat} P3HT, which competes with dissociation at donor-acceptor heterojunctions in photovoltaic diodes~\cite{Clarke2010}.

\begin{acknowledgments}
CS acknowledges NSERC and the Canada Research Chair in Organic Semiconductor Materials. GL and NS acknowledge the EC 7th Framework Program ONE-P project (Grant Agreement 212311). GL thanks the Royal Society. DB, Research Director of FNRS, is also partly supported by ONE-P (NMP3-LA-2008-212311), the Inter-university Attraction Pole IAP 6/27 of the Belgian Federal Government, and FNRS/FRFC. PLK acknowledges FQRNT, and MS acknowledges MELS/FQRNT.
\end{acknowledgments}


\begin{thebibliography}{28}
\bibitem{QuantumChem}
 Semi-empirical INDO/SCI excited-state calculations (including 200 molecular orbitals in the active space) have been performed on the basis of snapshots extracted from molecular dynamics simulations of the stack, using a force field parameterized against high-level ab-initio calculations. Further details will be reported elsewhere.
\end{thebibliography}

%

\end{document}